\pdfoutput=1
\documentclass{JINST}

\usepackage{listings, setspace,url}
\usepackage{verbatim}
\usepackage{dcolumn}

\usepackage{multirow}
\usepackage{xspace}
\usepackage{amsmath,mathtools}

\newcommand{\Fr}{$^{221}$Fr\xspace}
\newcommand{\Ac}{$^{225}$Ac\xspace}
\newcommand{\Ra}{$^{225}$Ra\xspace}

\newcommand{\At}{$^{217}$At\xspace}
\newcommand{\Bi}{$^{213}$Bi\xspace}
\newcommand{\Po}{$^{213}$Po\xspace}
\newcommand{\Tl}{$^{209}$Tl\xspace}
\newcommand{\Pb}{$^{209}$Pb\xspace}
\newcommand{\Frtwotwofive}{$^{225}$Fr\xspace}

\makeatletter
\renewcommand{\p@subsection}{}
\renewcommand{\p@subsubsection}{}
\makeatother

\newcommand{\arXivid}[1]{\href{http://www.arxiv.org/abs/#1}{\tt arXiv:#1}}

\newcommand{\tempnote}[1]{}  
\newcommand{\revnote}[1]{\xspace (\textbf{\emph{#1})}}

\newlength{\arrow}
\newcommand*{\myrightarrow}[2][]{\xrightarrow[#1]{\mathmakebox[\arrow]{#2}}}

\author{M.~Tandecki$^{a}$, J.~Zhang$^{b}$, S.~Aubin$^{c}$, J.~A.~Behr$^{a}$\thanks{Corresponding author, e-mail: behr@triumf.ca}, R.~Collister$^{d}$, E.~Gomez$^{e}$, G.~Gwinner$^{d}$, H.~Heggen$^{a}$, J.~Lassen$^{a}$, L.~A.~Orozco $^{b}$, M.~R.~Pearson$^{a}$, S.~Raeder$^{a}\thanks{Present address: IKS, KULeuven, 3000 Leuven, Belgium}$, A.~Teigelh{\"o}fer$^{a,d}$ \\
  \llap{$^a$}TRIUMF,   Vancouver, BC V6T 2A3, Canada,\\ \llap{$^b$}JQI, Department of Physics and NIST, University of Maryland,   College Park, MD 20742, USA,\\   \llap{$^c$}Department of Physics, College of William and Mary,   Williamsburg VA 2319, USA,\\ \llap{$^d$}Dept. of Physics and Astronomy, University of Manitoba,
  Winnipeg, MB R3T 2N2, Canada, \\  \llap{$^e$}Instituto de F{\'i}sica, Universidad Aut{\'o}noma de San Luis Potos{\'i}, San Luis Potos{\'i} 78290, Mexico  }

\abstract{We demonstrate a new technique to prepare an offline source of francium for trapping in a magneto-optical trap. Implanting a radioactive beam of \Ac, $t_{1/2} = 9.920(3)$\,days, in a foil, allows use of the decay products, i.e.~\Fr, $t_{1/2} = 288.0(4)$\,s. \Fr is ejected from the foil by the $\alpha$ decay of \Ac. This technique is compatible with the online accumulation of a laser-cooled atomic francium sample for a series of planned parity non-conservation measurements at TRIUMF.
We obtain a 34\,\% release efficiency for \Fr from the recoil source based on particle detector measurements. We find that laser cooling operation with the source is $8^{+10}_{-5}$ times less efficient than from a mass-separated ion beam of \Fr in the current geometry.
While the flux of this source is two to three orders of magnitude lower than typical francium beams from ISOL facilities, the source provides a longer-term supply of francium for offline studies.
}

\title{Offline trapping of \Fr in a magneto-optical trap from implantation of an \Ac ion beam}

\begin{document}

\tableofcontents

\section{Introduction}
\label{sec:introduction}

Trapped radioactive isotopes provide unique experimental systems for low-energy tests of the Standard Model \cite{behr09,severijns06}.
Recent and planned experiments include correlation measurements in nuclear $\beta$ decay \cite{gorelov05,scielzo04,flechard2011He,li2013tensor}, investigations of weak neutral currents \cite{gomez07,gomez06} and electric dipole moment (EDM) measurements \cite{parker2012efficient}.
The precision for these experiments requires high statistics, along with a detailed investigation of systematic effects. Radioactive ion beam (RIB) facilities or long-lived radioactive sources can provide the radioactive samples for these experiments.

RIB facilities can deliver a wide range of different isotopes; however, available beamtime is shared by many different experiments.
Long-lived radioactive sources complement -- or provide an alternative to -- radioactive ion beams.  They can be created in a short time ($ < 1$~day) during an online beamtime with a low chance of failure, while being able to deliver experimenters with a source that will last for weeks. Such a source can provide radioactive atoms for a longer duration which is crucial for experiments where systematic effects need to be investigated in detail. 
The drawbacks of using a radioactive source are its restriction to certain isotopes and additional radiological hazards.

Radioactive sources have been or are used in a number of experiments for testing fundamental symmetries, such as the EDM search in \Ra \cite{parker2012efficient}, trapping of \Fr \cite{lu97}, and correlation measurements in $\beta$ decay \cite{guckert98}, as well for nuclear structure studies using a $^{252}$Cf source \cite{savard2008radioactive}. In order to work with short-lived isotopes ($t_{1/2} < 1$~day), the source must include a long-lived precursor isotope. For example, the \Fr source of ref.~\cite{lu97} consists of an oven that releases the \Fr atoms into a magneto-optical trap (MOT), while retaining the long-lived \Ac precursor. In the case of the $^{82}$Rb source of \cite{guckert98}, the source consists of a $^{82}$Sr precursor which releases Rb ions, while the Sr remains bound in a molecule in the source; the $^{82}$Rb ions are subsequently mass separated to avoid $^{85}$Rb contamination. More recently, the CARIBU facility \cite{savard2008radioactive} has implemented a $^{252}$Cf precursor source, which uses an RF gas catcher to efficiently extract spontaneous fission decay products for use in experiments.

In this paper, we present the design, construction, and testing of a radioactive \Fr source based on a \Ac precursor sample implanted in a tantalum foil. The \Fr source was developed to support the program of weak interaction studies through parity non-conservation measurements in francium at the ISAC RIB facility at TRIUMF. The offline francium source is integrated into the trapping apparatus and can be used for testing the laser cooling system and investigating systematic effects in experiments, while the much higher yield francium beam produced directly by ISAC is essential to acquire statistics for precision measurements. Alternating between the two modes of operation is possible without opening the system to atmospheric pressure. The techniques mentioned in the previous paragraph do not allow this easily in our geometry, or they require a substantial modification of the infrastructure of ISAC. Two parity non-conservation experiments are planned; one to study the electric-dipole parity-forbidden $7S \rightarrow 8S$ transition and one to study the anapole moment using the microwave ground state hyperfine transition. For the $7S \rightarrow 8S$ experiment, \Fr can be used to optimize the system for the reduction of systematic effects \cite{wood99}. For the anapole experiment, \Fr will be hard to use for the final experiments, since its ground state splitting is significantly different from the neutron-deficient francium isotopes (i.e.~$15.5$\,GHz versus $\sim 45$\,GHz~\cite{duong87}).
Two examples of what can be done with this source for the latter project are the optimization of the dipole trap and measurement of differential AC Stark shifts \cite{sheng2013}, or the study of the N dependence of the signal from the atoms in a cavity (optical or microwave) to avoid effects associated with the weak coupling regime of cavity QED that may mix atomic levels in unwanted ways \cite{terraciano2007enhanced}.

The paper is organized as follows: the method is described, along with simulations, in section~\ref{sec:method}. We have performed several measurements to quantify the efficiency of the technique in section~\ref{sec:measurements}, which are discussed in section~\ref{sec:discussion}.

\section{Method}
\label{sec:method}

\begin{figure}
\begin{center}
  \includegraphics[width=0.95\linewidth]{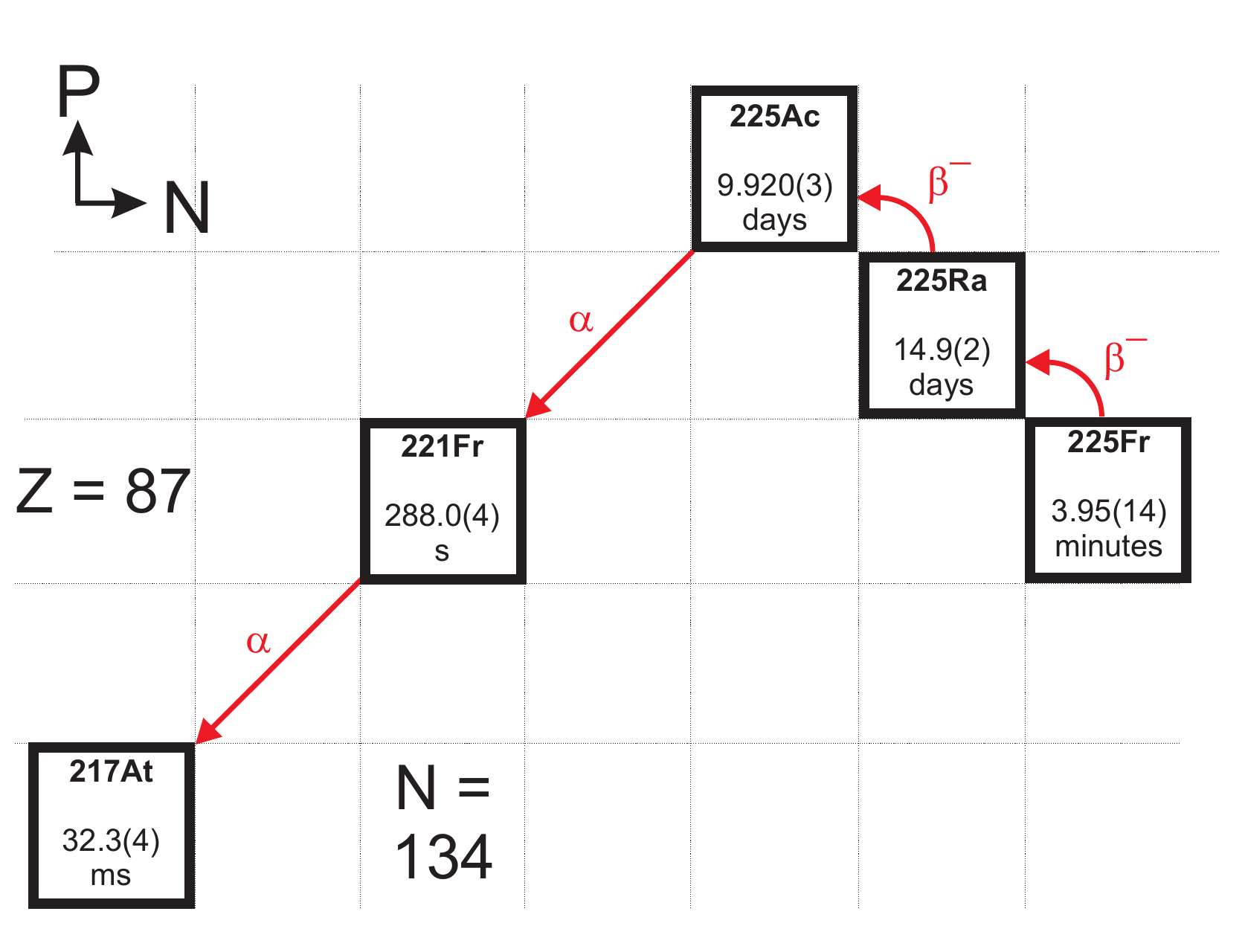}\\
  \caption{Decay chains for \Ac, \Ra and \Frtwotwofive; the proton number (P) is shown on the vertical axis and the neutron number (N) on the horizontal axis. The different boxes show the isotope on top and the corresponding half-life below. Branching ratios below 0.1\,\% are not listed here. This chain eventually ends with stable $^{209}$Bi.}\label{fig:ac_decay_chain}
  \end{center}
\end{figure}

The \Fr source consists of a precursor isotope, such as \Ac, implanted in a tantalum foil at a beam energy of $20$\,keV. When a \Ac nucleus undergoes alpha decay to \Fr, this daughter nucleus has a recoil energy of about $105$\,keV, which is sufficient to escape from the foil.  We use another foil made of yttrium, placed $3$\,mm away from the tantalum foil, to catch the recoiling \Fr ions and neutralize them; when the yttrium foil is heated it emits neutral \Fr atoms. The tantalum and yttrium foils are on independent mechanical stages so that they can be appropriately positioned for each of three operations: implantation, recoil catching, and neutralized emission into the MOT vapor cell.

As an alternative to implanting the source with the \Ac precursor, we can also use either \Ra or \Frtwotwofive. Both of these alternate precursor isotopes have a $\beta$ decay path to \Ac. Figure~\ref{fig:ac_decay_chain} shows all the relevant decay paths to $^{221}$Fr, and its subsequent $\alpha$ decay to $^{217}$At. The half-life of $^{221}$Fr is $t_{1/2}=288.0(4)$\,s, based on a weighted average of values found in \cite{jeppesen2007alpha,wauters10,suliman2013half}.

An \Ac source has a half-life of $9.920(3)$~days~\cite{pomme2012measurement}, and a \Fr rate which is initially $8.0 \cdot 10^{-7}$ of the implanted \Ac amount. A source consisting of \Ra has a half-life of $14.9(2)$~days, requiring a build-up time for the \Ra to decay to \Ac to reach a maximum \Fr rate of $2.4 \cdot 10^{-7}$ of the implanted \Ra after $17.5$~days. A \Frtwotwofive source is equivalent to one of \Ra for all intents and purposes, with its half-life of about $4$~minutes. A \Ra or \Frtwotwofive source results in a lower yield of \Fr, but the life of such a source is longer than an \Ac source. The isotope to choose depends on the experimental requirements and on the source isotope rate that can be delivered.

The \Fr rates from such a source should be compared to online francium yields which are of the order of $10^6$--$10^9$\,Fr/s for the neutron-deficient francium nuclei $^{206-213}$Fr at ISOL facilities \cite{isacyield,isoldeyield}. An implantation of \Frtwotwofive at maximum ISOLDE rates \cite{isoldeyield} can produce a source in little over $1$~hour which releases \Fr at rates of $1 \cdot 10^6$\,Fr/s; similarly, a $1$-day implantation results in a \Fr rate of $2 \cdot 10^7$\,\Fr/s. This is at the lower end of the range of online yields from UC$_x$ targets at ISOL facilities. However, the useful life of this source is longer than a month, whereas a francium source accumulated online is limited by the radioactive francium half-lives. 

The production of a source of \Ac at ISAC has one additional advantage compared to \Ra or \Frtwotwofive; after a long-term exposure to proton irradiation of a UC$_x$ target, \Ac continues to be released from the target if it remains hot ($\sim 2200^\circ$\,C), whereas \Ra diffuses out of the target within hours. The amount of \Ac that stays in the target will be higher for thick sources (i.e.~at ISOLDE), whereas at ISAC recent UC$_x$ targets consist of thin wafers to enhance the diffusion of short-lived species like francium. After an online running period, targets are typically not removed immediately to let them `cool down' for some period.  It is possible to create the required source without interfering with online beamtimes and minimal interference with target maintenance operations. This is how the \Ac source was created for the experiment described in this paper.



The integration of the \Fr source into the francium trapping apparatus is explained in section~\ref{sec:setup}, while efficiencies and limitations of the technique for our geometry, isotope and implantation energy are explored through simulations in section~\ref{sec:SRIM}. Other possibilities using the same technique are briefly noted in section~\ref{sec:other_applications}.

\subsection{Experimental setup}
\label{sec:setup}


\begin{figure}
\begin{center}
  \includegraphics[width=\linewidth]{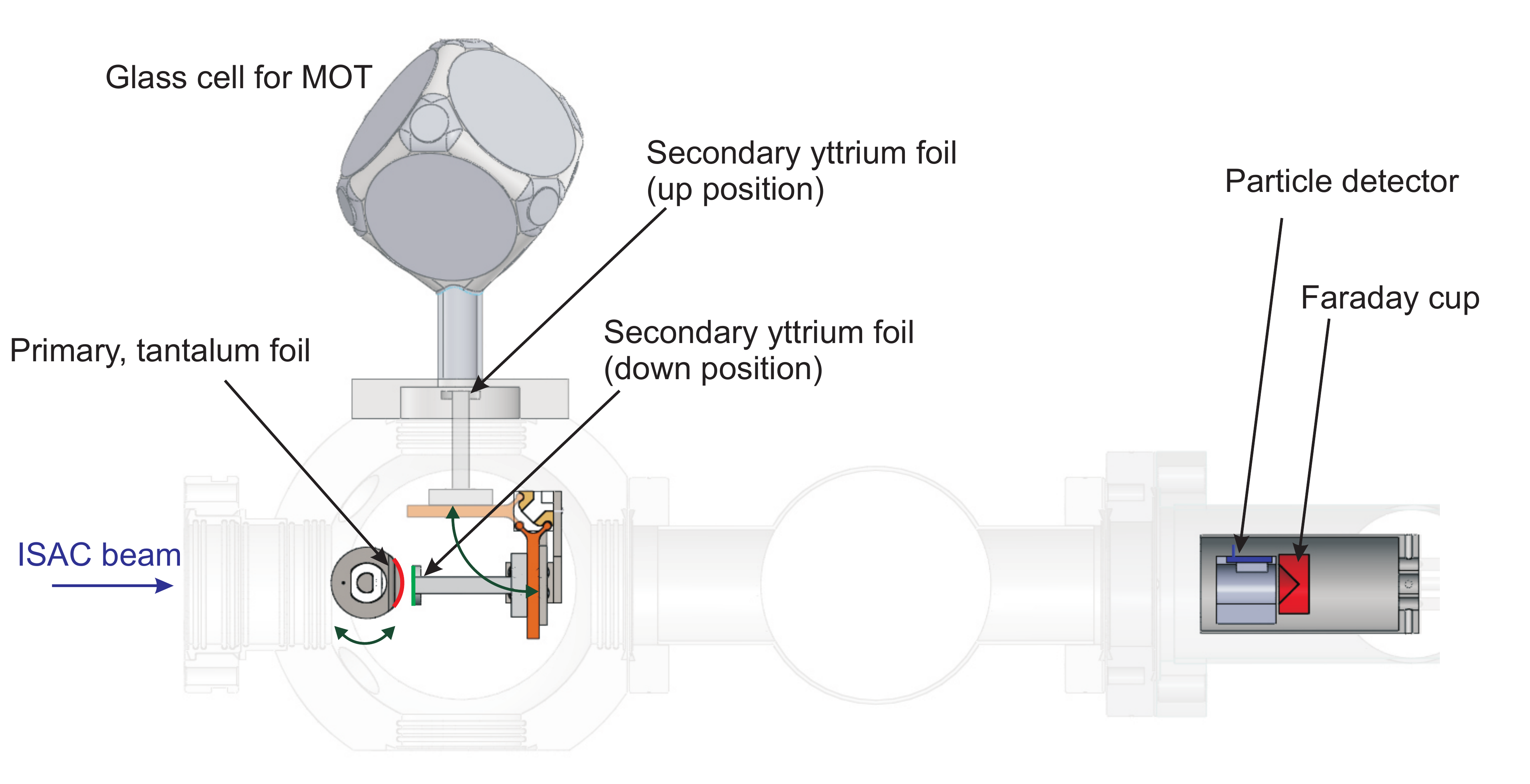}\\
  \caption{Schematic figure of the apparatus, highlighting parts relevant fort his paper: shown on the left, the yttrium (secondary) foil -- which rotates upwards to release francium in the glass cell -- and the tantalum (primary) foil, and shown on the right, the particle detector in blue and the Faraday cup in red.}\label{fig:alpha_faraday_cup}\label{fig:exp_setup}
  \end{center}
\end{figure}

Ref.~\cite{tandecki2013commissioning} explains in detail the online operation of the francium trapping apparatus. Briefly, among many other elements, francium is created at the ISAC facility at TRIUMF (see~\cite{dombsky00}) by bombarding a UC$_x$ target with $500$\,MeV protons with an intensity up to $10$\,$\mu$A for this type of target. The francium atoms are surface-ionized, pass through a mass separator, and are delivered to the experiment with a beam energy of typically $20$--$30$\,keV. In our experiment the ion beam is focussed onto an yttrium foil that accumulates and neutralizes the francium ions. We will refer to this yttrium catcher foil as the secondary foil in this paper. After the accumulation, typically $20$\,s, the foil rotates $90^\circ$ upwards out of the path of the ion beam and faces a glass cell. Then it is heated by running a current through it to release neutral francium.
The francium atoms are released into the glass cell and collected in a MOT, which is a combination of three perpendicular pairs of counter-propagating laser beams and a quadrupole magnetic field. 

To add the functionality of the offline \Fr source a tantalum foil is added to the system upstream of the yttrium foil, as shown in figure~\ref{fig:exp_setup}.  Tantalum was chosen because it is an inert material. We will refer to this foil as the primary foil. The tantalum foil itself is mounted on a cylindrical holder on a rotational feedthrough, which allows it to face the ion beam, to face the yttrium foil and to be outside of the path of the ion beam for normal online operation. When they are facing each other, the gap between the primary and secondary foils is $\sim 3$\,mm. For this measurement the yttrium foil was circular with a $1$\,cm diameter, while the tantalum foil is somewhat larger than this. An \Ac beam from ISAC impinging on the primary foil generates the \Ac source. After adequate accumulation, we can rotate the primary foil to face the secondary foil so that \Fr can be implanted in the latter. Each of these two assemblies is mounted independently on its own CF$4.5$'' flange to enable the possibility of easy removal or exchange to limit radiation exposure.

\subsection{Calculations of collection efficiencies}
\label{sec:SRIM}

\tempnote{Details of the SRIM simulations can be found in appendix \ref{app:srim}.}

\begin{figure}[!thb]
\begin{center}
  \includegraphics[width=0.95\linewidth]{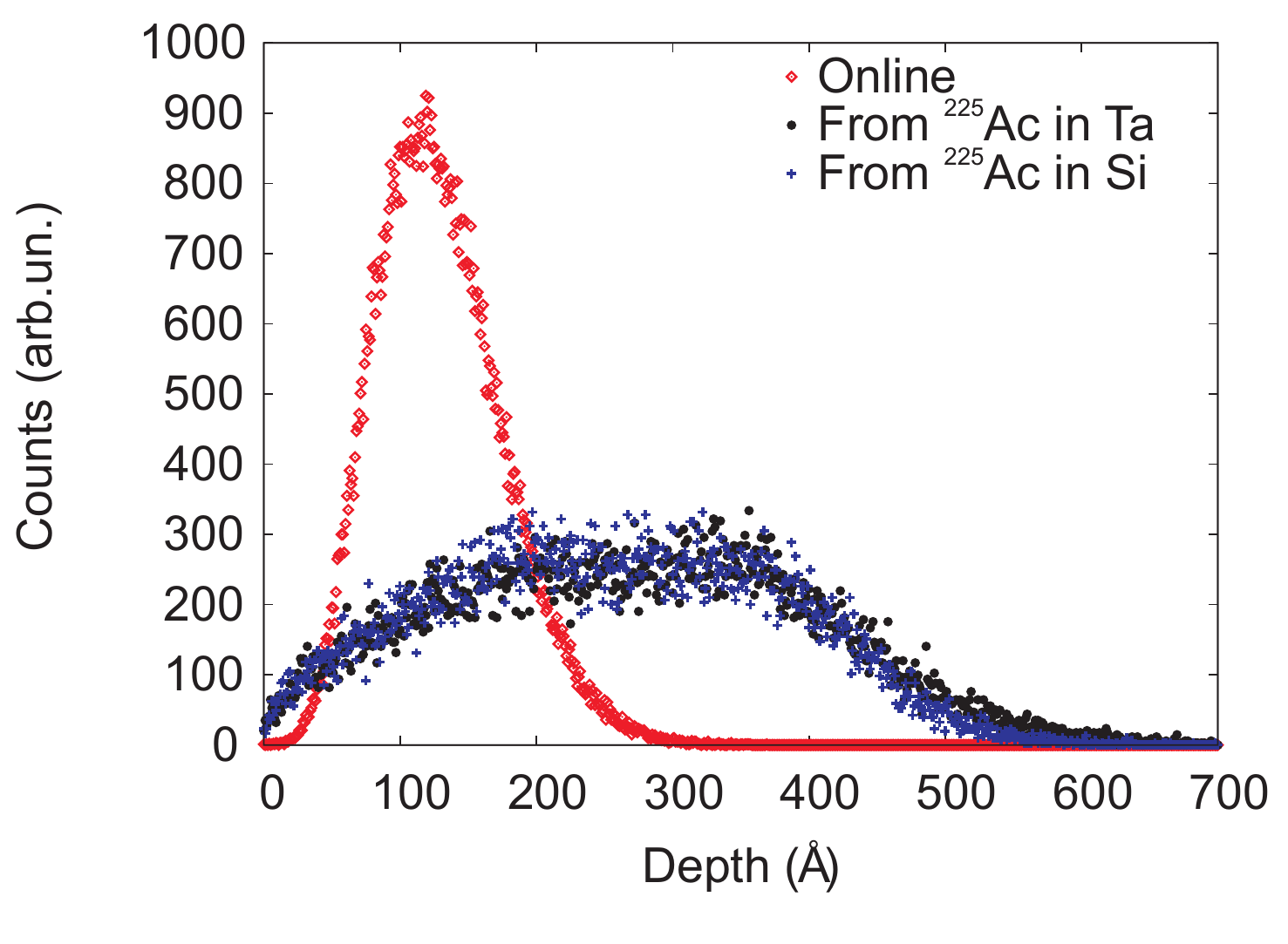}\\
  \caption{Implantation profiles for \Fr into Y from different sources. In (red) diamonds \Fr online from ISAC, in (blue) crosses \Fr from \Ac in a Ta foil and in (black) dots \Fr from \Ac in a Si foil. The data are normalized such that the integrated count rate is one million for each curve.   }\label{fig:Y_implantation}
  \end{center}
\end{figure}

\begin{figure}
\begin{center}
  \includegraphics[width=0.80\linewidth]{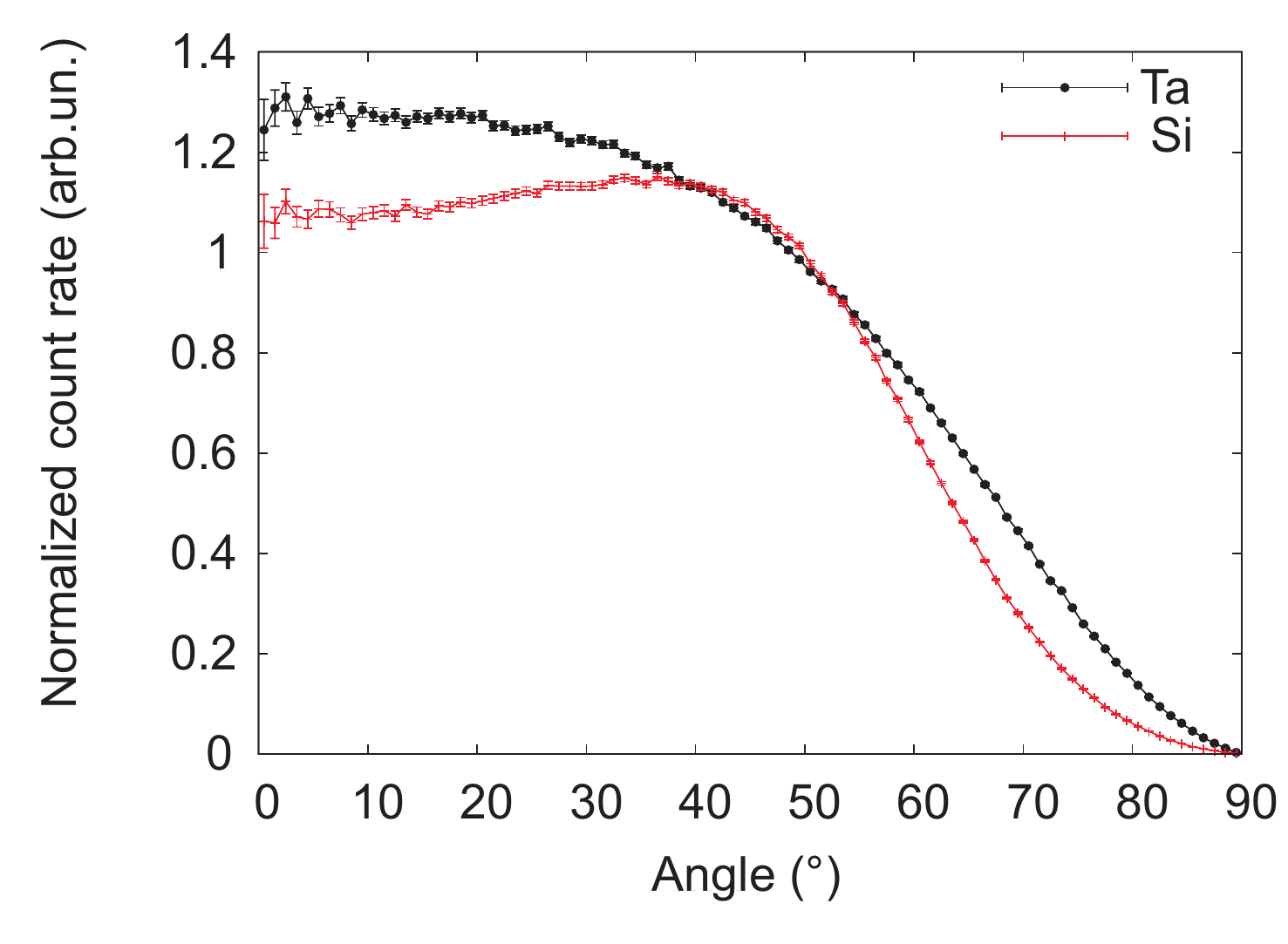}\\
  \caption{Histogram (with bins of $1^{\circ}$) showing the total emission rate of \Fr out of tantalum (black line with dots) and silicon (red line without dots), as a function of the emission angle with respect to the surface normal. Data are normalized to the angular emission rate for isotropic emission over $4 \pi$.  }\label{fig:implantation_efficiency}
  \end{center}
\end{figure}

The goal of the simulations is to estimate the efficiency of the technique; this was done with the SRIM simulation code \cite{ziegler2010srim}.
The \Ac implantation energy is $20$\,keV, the primary foil material is tantalum and the geometry is as described above (with a distance of $3$\,mm between the two foils being the most important parameter).
The simulations are performed as realistically as possible within the SRIM simulation environment.
We start with a $20$\,keV \Ac beam implanted into a tantalum foil; the resulting implantation profile, which has an average implantation depth of $55$\,{\AA} with a straggle of $22$\,{\AA}, is used as input for a second simulation to characterize the transport of recoiling francium from the first foil into the second one.
The online beam is modelled as coming from one point with zero emittance; realistic beam profiles can be obtained by convolution. The curvature of the primary foil (over the beam spot) is small enough to introduce only small corrections.
Figure~\ref{fig:Y_implantation} shows implantation profiles into yttrium of a $20$\,keV \Fr beam and of \Fr emitted from a tantalum and silicon foil. The implantation depth from the online beam is $129$\,{\AA} with a straggle of $46$\,{\AA}. The depth from the offline sources is on average more ($270$\,{\AA} for Ta), owing to the higher energy of $\sim 105$\,keV minus $20$\,keV, but a fraction is also implanted close to the surface; the \Fr decay products emanate from a localized source and recoil at random angles, whereas the online source comes from a parallel beam.
Different primary foil materials result in slightly different emission efficiencies. Tantalum ($Z = 73$, $A = 181$), a relatively heavy nucleus, causes a lot of energy straggling of the implanted beam. Silicon ($Z = 14$, $A = 28$--$30$) on the other hand, despite causing less straggling, does not result in a higher efficiency, $31$\,\%, compared with $34$\,\% for tantalum. 
Figure~\ref{fig:implantation_efficiency} shows the emission rate as a function of the emission angle out of the foil. 
Surprisingly, at small angles, the rate is higher than expected; atoms emitted at larger angles, and initially backscattered atoms as well, contribute to this higher rate at small angles, while reducing the rate at larger angles.

A side effect of this point-like source is that the lateral distribution on the yttrium foil will be larger than from an online beam; an \Ac beam having a Gaussian lateral profile with a 1$\sigma$ radius of $1.5$\,mm, or $90$\,\% within a $3.2$\,mm radius, with a distance of $3$\,mm ($5$\,mm) between the two foils, results in a \Fr distribution with a Gaussian profile with a 1$\sigma$ radius of $3$\,mm ($4.5$\,mm). Therefore, $75$\,\% ($46$\,\%) ends up on the $1$\,cm-diameter foil and $45$\,\% ($22$\,\%) in the same $3.2$\,mm radius as the mass-separated beam.

These two effects, a larger average implantation depth and a larger spot size, have an important effect for our experiments. In order to load the MOT, we dispense the \Fr embedded in the yttrium foil by heating it. The heating is done by sending a current through the foil (typically $\sim$\,$10$\,A for $\sim$\,$1$\,s).
The deeper average implantation of the \Fr in the yttrium reduces the dispensed amount roughly by a factor of two. This is estimated by using the diffusion coefficient of francium in yttrium, $1.6(9) \cdot 10^{-14}$ \cite{demauro2008}, and equation 3 in reference \cite{melconian05} for the time-dependent released fraction.
This loss can be minimized by increasing the amount or duration of the heating, which was not done at this time to avoid possible damage to the foil.
Visual inspection shows that the heating profile is a stripe with a larger temperature in the center and a lower temperature towards the legs, where the heat sinks. The area that is effectively heated is about $1/3$ of the total foil size. Since the spot size from \Fr escaping the tantalum foil is much larger than the online beam spot size, one expects -- for the same heating conditions -- a decrease in efficiency. Also, the $1$\,cm-diameter neck of the glass cell for the MOT limits the amount of \Fr that can be trapped in the MOT (see figure~\ref{fig:exp_setup}).

\subsection{Other isotopes}
\label{sec:other_applications}

Extensions to other isotopes and elements are straightforward. The only requirement for the source is that the mother isotope of the isotope of interest should undergo $\alpha$ decay (giving the daughter nucleus a sufficiently high recoil energy), and the initial source should be reasonably long-lived, longer than a week. We are interested in \Fr, but \At is released as well into the yttrium foil with an efficiency of $12$\,\%. 

Finally, a different isotope of francium, $^{223}$Fr, could be used in our geometry by forming an $^{227}$Ac source. However, the $\alpha$ branching ratio of $^{227}$Ac is low ($1.4$\,\%) so that a sizeable source would be required to generate useable quantities of $^{223}$Fr, presenting a significant radiological hazard. This decay has been used in the past to study the structure of $^{223}$Fr by $\alpha$ and $\gamma$ spectroscopy~\cite{sheline1995experimental}.



\section{Experiment}
\label{sec:measurements}

A  $20$\,keV $^{225}$Ac+ beam impinges on the primary foil for $18.5(5)$\,hours of at an average rate of $6.0(6)$\,pA (or $3.7(4) \cdot 10^7$\,Ac/s) from the ISAC target without active proton irradiation -- but after an online running period.
In-source resonant laser ionization using a two-step laser excitation scheme into an auto-ionizing state in \Ac \cite{lassen2006resonant,raeder2013source} enhances the extraction efficiency by two orders of magnitude over pure surface ionization.
Earlier tests on a UC$_x$ target showed that during proton irradiation one third of produced isotopes at mass 225 is $^{225}$Ac, while the rest is mainly $^{225}$Ra. Thanks to the in-source resonant laser ionization any contribution from contaminants is negligible. A total of $2.5(3) \cdot 10^{12}$ \Ac atoms are implanted in the foil. \tempnote{Calculation of number + error bars in appendix \ref{app:fr_implantation}} Using the \Ac half-life of $9.920$\,days yields a total decay rate of \Ac into \Fr of $1.9(2) \cdot 10^6$/s, immediately after the end of the implantation.

\subsection{Particle detector measurement}

The \Ac source rotates towards the Faraday cup and $\alpha$ detector assembly (see figure~\ref{fig:alpha_faraday_cup}) and the yttrium foil rotates out of the way to measure the \Fr release rate for comparison with simulations.
The solid angle of the Faraday cup seen from the \Ac source is $0.045(2)$\,\%, and the solid angle of the particle detector as seen from the center of the Faraday cup is $1.4(1)$\,\%\tempnote{see appendix~\ref{app:solid_angles}}. As the \Ac source faces the Faraday cup a source of \Fr builds up on the Faraday cup.
The particle detector is a silicon photodiode (Hamamatsu S3590-09).
The signal processing consists of a shaping amplifier, a discriminator and a multichannel scaler.
Therefore, no energy discrimination of the detected particles is done.

The decay chain starting at \Fr contains many $\alpha$ and/or $\beta$ decaying isotopes (see e.g.~\cite{suliman2013half} including half-lives): \Fr, \At, \Bi, \Po, \Tl, and \Pb.
Isotopes with a short half-life ($< 1$\,s) will be detected with the same effective half-life as the mother isotope: \At together with \Fr and \Po, \Tl together with \Bi.
Isotopes with a half-life longer than the time interval used for the analysis (i.e.~$1000$\,s) will not influence the analysis: \Pb.
The remaining relevant quantities are therefore the half-lives of \Fr and \Bi, $288.0(4)$\,s and $2737(4)$\,s respectively.

\begin{figure}[!thb]
\begin{center}
  \includegraphics[width=0.95\linewidth]{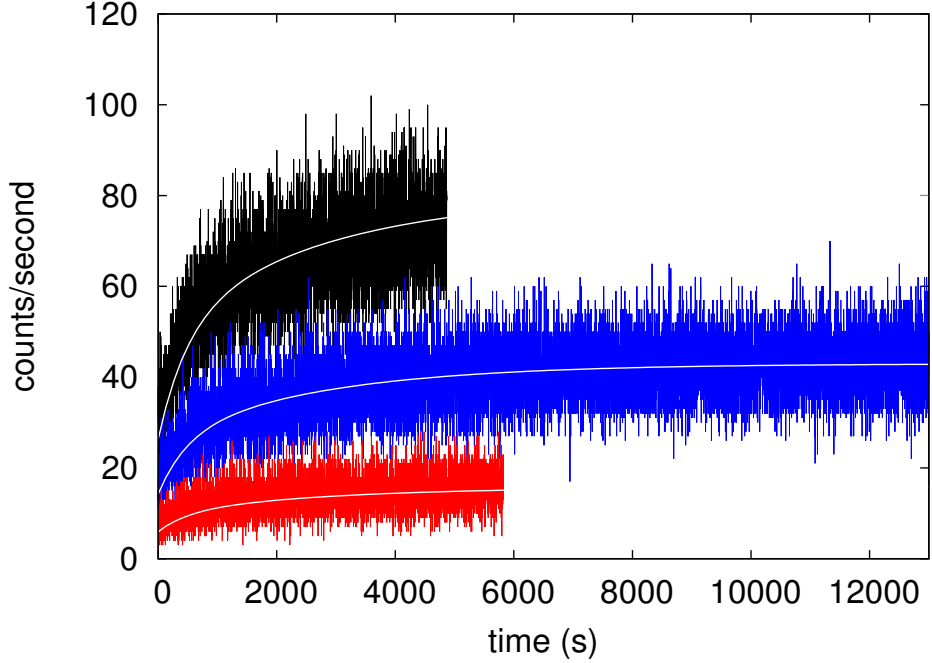}\\
  \caption{Buildup of a \Fr source on the Faraday cup as detected by the particle detector, with $t=0$\,s being the start of the \Fr implantation: 2 hours (black), 9 days (blue) and 23 days (red) after the \Ac implantation. The white lines overlaid on the data are the best fits for the data sets.}\label{fig:activity_buildup}
  \end{center}
\end{figure}


\newcommand{\tFr}{\tau_{\mbox{\tiny{Fr}}}\xspace}
\newcommand{\tBi}{\tau_{\mbox{\tiny{Bi}}}\xspace}

We have made measurements with the particle detector of the \Fr produced by the  \Ac source at three different times: two hours after the end of the \Ac implantation and then $9$ and $23$~days later.
The data are shown in figure~\ref{fig:activity_buildup}.
A simple model of the growth of the \Fr population deposited in the Faraday cup at a constant rate, where the activity decays into one daughter (\Bi) that in turn decays as well, explains the data reasonably well, as shown in table~\ref{tab:fit_results}. The count rate on the particle detector as a function of time, $R_{\mbox{total}}(t)$, is then:
\begin{equation}
\begin{split}
R_{\mbox{total}}(t) = {} & B + R_{\mbox{\tiny{Fr}}} \cdot \Big[ 1-exp(-t/\tFr ) \Big] \\
 & +  \frac{  R_{\mbox{\tiny{Fr}}}}{ \tFr - \tBi } \cdot \Big[ \tFr \cdot \big( 1 - exp(-t/\tFr) \big) \\
 & - \tBi \cdot \big( 1-exp(-t/\tBi ) \big) \Big],
 \end{split}\label{eq:rate}
\end{equation}
with $\tFr$ and $\tBi$ the \Fr and \Bi lifetimes, respectively, $B$ the background count rate and $R_{\mbox{\tiny{Fr}}}$ the \Fr implantation rate. 
The background count rate comes from activity present before starting the \Fr implantation, while the \Fr and \Bi half-lives are well-known quantities. We validate the model by leaving the \Fr and \Bi lifetimes free, but once validated, the model only has the \Fr implantation rate as free parameter.


\begin{table}[h]
\caption{Fit results for the data from the particle detector measurements. $\chi^2$/NDF is the reduced chi-squared.   }
\begin{center}
\begin{tabular}{|r c c|}
\hline
Days after     &  \Fr rate (s$^{-1}$)        & $\chi^2$/NDF  \\
 implantation    &             &   \\\hline
0.083 &    29.6(3)\phantom{00}       &  1031\,/\,947 \\
9\phantom{.083} &   15.2(2)        &  1080\,/\,999  \\
23\phantom{.083} &    \phantom{1}4.83(14)       &   1028\,/\,999  \\ \hline
\end{tabular}
\end{center}
\label{tab:fit_results}
\end{table}

For the final analysis we used the first $1000$\,s of each data set, since we are interested in the \Fr rate and we want to avoid sensitivity to activity migrating in the system.
Fitting an exponential decay through the three  \Fr rates gives $30.0(6)$\,Fr/s, right after the \Ac implantation. The fitted lifetime is consistent with that of \Ac decay, indicating the \Fr source in the tantalum foil has not decayed by other means, e.g.~by diffusion out of the primary foil or by having another element present in the system associated with the trapping facility (e.g.~Rb, see Ref.~\cite{tandecki2013commissioning}). 

Starting from the estimated \Fr rate escaping the foil, $1.9(2) \cdot 10^6$~\Fr/s (from measured \Ac rates during the implantation), and taking into account solid angles of $0.045(2)$\,\% (\Fr source to Faraday cup) and $1.4(1)$\,\% (Faraday cup to particle detector), an increased \Fr emission rate for these angles by a factor of $1.28(3)$ (see figure~\ref{fig:implantation_efficiency}), double counting because of \At decay, leads to an expected rate on the particle detector of $31(4)$\,counts/s. This number is in good agreement with the rates of $30.0(6)$\,counts/s deduced from particle detector measurements.

\subsection{Magneto-optical trap measurements}

The final test of our offline \Fr source is to verify that we can use it to load \Fr into a MOT. We operate the MOT with $120$\,mW (total over 6 beams) of trapping light on the D2 cycling transition ($718$\,nm) and $8$\,mW of repumper light on the D1 transition ($817$\,nm). The trapping beams each have an area of about $\sim 20$\,cm$^2$. The magnetic gradient of the MOT is 7 G/cm along the strong axis.
The correct trapping and repumping laser frequencies were approximatively maintained from the online run one week before through the transfer cavity lock \cite{zhao98}. \Fr was implanted into the yttrium foil for $10$~minutes, about one week ($177$~hours) after the implantation of the \Ac. Starting from a \Fr rate of $1.9(2) \cdot 10^6$/s, this yields $1.78(19) \cdot 10^8$ \Fr implanted in the yttrium foil and available for trapping. \tempnote{See appendix~\ref{app:number_fr_trapping} for the details} From these a fraction of $1.3(6) \cdot 10^{-5}$ was trapped in the MOT, with a total trapped amount of $2300(900)$.
This was measured by the fluorescence of the atoms and the total uncertainty is dominated by the uncertainty in total laser power at the atom cloud position and the amount of detuning from resonance.
The efficiency during the online  run of $5(2) \cdot 10^{-4}$ comes from trapping $^{209}$Fr. For \Fr the efficiency was measured to be a factor of $5$ lower during the online run; this can be understood qualitatively from their different atomic structures. The \Fr atom has a relatively narrow hyperfine splitting, $58$\,MHz between the cycling trapping transition ($7S_{1/2} F= 3 \rightarrow 7P_{3/2} F = 4$) and a transition that can decay to the lower hyperfine ground state level ($7S_{1/2} F= 3 \rightarrow 7P_{3/2} F = 3$), as compared to $518$\,MHz for $^{209}$Fr. A simple 6-level rate equation calculation suggests that for our repumper parameters only $10$--$15$\,\% is in the $7P_{3/2}$ level at any given time, which reduces the fluorescence per trapped atom proportionally and also reduces the collection efficiency of atoms into the MOT. The process of trapping francium from the online ion beam as compared to \Fr from the offline \Ac source is $8_{-5}^{+10}$ times more efficient, normalized to the amount of \Fr implanted in the secondary foil. 
Due to the long implantation times required to accumulate a sufficient amount of \Fr into the yttrium foil, no systematic studies using the \Ac source were performed, because at the time e.g.~the long-term laser stability was not good enough yet.

In the optimistic, $1 \sigma$-lower-limit case the efficiency of the offline scheme is only $3$ times less efficient that the online mode of operation in our measurement. This can be explained by (i) having a non-optimised MOT, or factors inherent to our geometry such as (ii) the more diffuse source of \Fr on the yttrium foil and (iii) the deeper implantation profile. In the pessimistic, $1 \sigma$-upper-limit case one can add that there were some mechanical problems in the system, before this test, causing the yttrium foil to be slightly damaged. This resulted in an irregular heating profile. From our measurement, we are unable to differentiate between losses from each of those separate effects; they are listed in table~\ref{tab:efficiency} with an estimate of their effect.

\begin{table*}[!ht]
\caption{Summary of losses from \Fr escaping the tantalum towards trapping in a MOT. The losses are divided between losses that are inherent to our apparatus (`overall'), and losses which are only applicable to the technique described in this paper (`offline').    }
\begin{center}
\begin{tabular}{| c c | r |  l |}
\hline
\multicolumn{2}{| c | }{Loss factor} & \Fr amount & Explanation of loss factor \\
overall & source &  &  \\
\hline
 &  & $5.20 \cdot 10^8$ &	\Fr amount in the primary foil  \\
 & $3$    &   $1.78 \cdot 10^8$   & Francium entering the secondary foil \\
$2000$  &      &   $8.9 \cdot 10^4$    & Online trapping efficiency of $^{209}$Fr \\
$5$ &     &  $1.78 \cdot 10^4$       &   \Fr trapping efficiency as compared to $^{209}$Fr  \\
 & $2$--$5$    &         &  Implantation depth  \\
 & $2$--$5$     &         &  Larger \Fr spot size  \\
  & $1$--$10$     &         &  Possible damage to yttrium foil  \\
  & & $2.3 \cdot 10^3$ &	Trapped francium\\
\hline
\end{tabular}
\end{center}
\label{tab:efficiency}
\end{table*}

\section{Discussion}
\label{sec:discussion}

The $\alpha$ detection measurements suggest no fundamental limitation to this technique; $34$\,\% of implanted \Ac will escape the foil as \Fr for our experimental conditions. Our geometry is optimised for online implantation of francium isotopes for this first measurement of trapping \Fr from an \Ac source. In the future a better geometry could be found; the addition of a translational feedthrough, for instance, would minimize the distance between the two foils avoiding losses from having a larger francium implantation spot size. The larger implantation depth could be compensated for by an improved operation of the yttrium foil, e.g.~more intense heating. The efficiency achieved can now be used to investigate systematic effects (such as those in section~\ref{sec:introduction}).


The overall efficiency of our technique is comparable to the experiments mentioned in the introduction. The \Ac source of \cite{lu97} had an efficiency of $2$\,\% for \Fr leaving the oven, the $^{82}$Sr source of \cite{guckert98} obtained $35$\,\%, and the CARIBU facility in \cite{savard2008radioactive} achieves $35$\,\% for incidence rates of the order of $10^6$\,atoms/s. The latter two experiments have the advantage of extracting an ion beam, allowing for more favorable implantation properties on the catcher foil. The required infrastructure, however, is much more complex than the one used in our method. While the first method of generating \Fr in an oven  can be implemented in a relatively small space, it does not allow both online and offline francium trapping in our geometry.

A key component of our technique is that it permits the use of the \Ac source in combination with online beamtime, without modifying the setup and without requiring a modification of the ISAC beamlines further upstream.
This can extend the available development time with \Fr, while allowing high-statistics measurements with online beamtime. Eventually this dual use could be superseded by longer beamtimes becoming available, as promised by the ARIEL facility \cite{dilling2014ariel}, under construction at TRIUMF, or the planned ISOL@MYRRHA facility \cite{nilsson2013european}.

\section{Conclusion}
\label{sec:conclusion}


We have demonstrated magneto-optical trapping of \Fr ejected by
$\alpha$ decays from a \Ac source made by implanting a mass-separated
ion beam. The use of ejected \Fr provides one solution to technical difficulties encountered by other methods when trapping decay products
released by heating generator material (see Ref. \cite{lu97} and \cite{guckert98}). Such an \Ac source foil made by implantation could in principle be transportable to remote locations where the primary beam is not available, as can be done at ISOL facilities.



\acknowledgments

This work is supported by NSERC and NRC from Canada, NSF and DOE from USA, CONACYT from Mexico. We acknowledge helpful discussions with H.~Gould, K.~Johnston, M.~Kowalska, P.~Kunz and M.~Lindroos. 

\end{document}